\documentstyle [12pt]{article}
\newcommand{\be}{\begin{equation}}
\newcommand{\ee}{\end{equation}}
\newcommand{\bees}{\begin{eqnarray}}
\newcommand{\ees}{\end{eqnarray}}
\newcommand{\ra}{\rightarrow}

\newcommand{\lsim}{\stackrel{<}{\sim}}

\newcommand{\vr}{\vrule width 0pt height 15pt depth 15pt}
\newcommand{\vrbig}{\vrule width 0pt height 27pt depth 15pt}

\begin{document}

\begin{titlepage}
\begin{flushright}
 IFUP-TH 25/96\\
  May   1996\\
  gr-qc/9605072
\end{flushright}

\vspace{5mm}

\begin{center}

{\large\bf  Spectrum of relic gravitational waves\\

\vspace{4mm}

in string cosmology}

\vspace{10mm}

{\large Alessandra Buonanno, Michele~Maggiore}
and  {\large Carlo~Ungarelli}

\vspace{3mm}

Dipartimento di Fisica dell'Universit\`{a} and INFN,\\
piazza Torricelli 2, I-56100 Pisa, Italy.\\

\end{center}

\vspace{4mm}

\begin{quote}
\hspace*{5mm} {\bf Abstract.} We compute  the spectrum of relic
gravitons in a model of string cosmology. 
In the low- and in the high-frequency limits we reproduce  known
results. The full spectrum, however, also displays a  series of
oscillations which could give a characteristic signature 
at the planned
LIGO/VIRGO detectors. For special values of the parameters of the
model the signal reaches its maximum already at frequencies
accessible to LIGO and VIRGO and it   is close to the sensitivity of
first generation experiments.

\end{quote}
\end{titlepage}
\clearpage

\section{Introduction}
In the next few years a number of detectors for gravitational waves,
and in particular the LIGO and VIRGO interferometers, are expected to
start operating in a range of frequencies between  $10\, {\rm Hz}$ and 
$1\, {\rm kHz}$. One of the possible signals which could be searched,
correlating the output of two detectors, 
is a stochastic background of gravitational waves.
This background is expected to have different components, with
different origin: it will get contributions from a large number of
unresolved sources at modest red-shifts, as well as from radiation of
cosmological origin. The latter is especially interesting, since it
would carry  informations about the state of the very
early universe.

The basic mechanism of generation of relic gravitational waves in
cosmology has been discussed in a number of papers, 
see e.g. refs.~\cite{Gri,vari}, the reviews~\cite{Gri2,Muk} and
references therein. The spectrum can be conveniently expressed using
$$
\Omega_{\rm gw}(f)=\frac{1}{\rho_c}\,\frac{d\rho_{\rm gw}}{d\log f}
$$
where $\rho_c$ is the critical density of the universe,
$\rho_{\rm gw}$ is the energy density in gravitational waves and $f$ 
is the frequency. 
Particular attention has been paid to the spectrum produced in
inflationary cosmology. In this case one finds that
$\Omega_{\rm gw}$
decreases with frequency 
as $f^{-2}$ from $10^{-18}\,{\rm Hz}$ to $10^{-16}\,{\rm Hz}$, and
then it is flat up to a maximum cutoff 
frequency $f\sim 10^{10}\,{\rm Hz}$. 
While the frequency dependence of $\Omega_{\rm gw}(f)$ is fixed, its
magnitude depends on a parameter of the model, the Hubble constant
during inflation. An upper bound on the spectrum 
can be obtained from the
measurement of COBE of the anisotropy of the microwave background
radiation. Via the Sachs-Wolfe effect, a large energy density in
gravitational waves at wavelengths comparable to the present Hubble
radius would produce fluctuations in the temperature of the photon 
cosmic background. This gives a limit 
on $\Omega_{\rm gw}$~\cite{Allenrev}
of about $8\cdot 10^{-14}$ at $f\sim 10^{-16}\,{\rm Hz}$ . Since for
larger frequencies  the spectrum predicted by inflation is flat, this
bound also holds at the frequencies of interest for LIGO and VIRGO. 
The planned
sensitivities of these experiments to a stochastic background are
of the order of $\Omega_{\rm gw}\sim 5\cdot 10^{-6}$, 
while the advanced LIGO
project aims at $5\cdot 10^{-11}$~\cite{Allenrev}.
 In any case, the spectrum predicted
by these inflationary models is too low to be observed.

Clearly, in order to have a stochastic background which 
satisfies the COBE
bound, but still has a chance of being observable at LIGO or VIRGO,
the spectrum must grow with frequency. A spectrum of this type has
been found in ref.~\cite{BGGV} in a  cosmological model suggested by
string theory~\cite{GV,review,review2}.  
Because of its fast ( $\sim f^3$) growth
with frequency at low $f$, the COBE bound is easily evaded, and the
most relevant bound for this type of spectrum comes from
nucleosynthesis. The result is that, for a  certain range of
values of the parameters of the model, the spectrum might be 
accessible at
the  interferometer experiments, at least at the advanced level, while
satisfying the existing experimental bounds.

In ref.~\cite{BGGV} this spectrum has been estimated, using approximate
methods, in the low- and in the high-frequency limits, and neglecting
overall numerical factors. In this paper
we present a detailed  computation of the spectrum, 
solving exactly the relevant differential equations. We fix
the numerical factors and we present the frequency dependence in the
intermediate region. The latter displays an interesting feature:
it shows a series of 
oscillations, which might provide a characteristic experimental
signature. 

As remarked in~\cite{BGGV}, one must be aware of the fact
that it might not be legitimate to use field-theoretical methods
during the 'stringy phase' of the cosmological model, see sect.~2, and
large frequencies are indeed sensitive to this phase.
However the best one can do, at this stage,
is to write down a specific cosmological model and see what are its
predictions. Of course, these predictions should only be considered as
indicative.

\section{The model}
The low energy string effective action depends on the metric
 $g_{\mu\nu}$ and on the dilaton field $\phi$ (we neglect the
 antisymmetric tensor field).
At lowest order in the derivatives and in $e^{\phi}$ it is given by
\be\label{Seff}
S=-\frac{1}{2\lambda_s^2}\int d^4x\, \sqrt{-g}\, \left [ e^{-\phi}
\left( R+\partial_{\mu}\phi\partial^{\mu}\phi\right)
 -V_{\rm dil}(\phi )\right ]\, ,
\ee
where $\lambda_s$ is the string lenght 
and $\phi$ is the dilaton field. The 
dilaton potential $V_{\rm dil}(\phi )$ is due to
 non-perturbative effects and therefore vanishes  as 
$\exp (-{ c}\exp (-\phi ))$ for $\phi$ large and negative, with
$c$ a positive constant.
We consider a homogeneous, isotropic and spatially flat background,
$\phi =\phi (t), ds^2=dt^2-a^2(t)\,d{\bf x}^2$,
and we introduce conformal time $\eta$, $dt =a(\eta )\,d\eta$. The
pre-big-bang scenario proposed by Gasperini and 
Veneziano~\cite{GV,review,review2}
suggests the following choice for the background metric and dilaton 
field.

For $-\infty <\eta <\eta_s$, with $\eta_s <0$, we have
 a dilaton-dominated  regime with
\bees\label{dil}
a(\eta )&=&-\frac{1}{H_s\eta_s}\left( 
\frac{\eta-(1-\alpha )\eta_s}{\alpha\eta_s}\right)^{-\alpha}\\
\phi (\eta )&=&\phi_s-\gamma\log
\frac{\eta-(1-\alpha )\eta_s}{\eta_s}\, .
\ees
With the values $\alpha = 1/(1+\sqrt{3}),\gamma =\sqrt{3}$ this is a
solution of the  equations of motion derived from the
effective action~(\ref{Seff}) in the absence of
external matter~\cite{GV}.

At a value $\eta =\eta_s$ the curvature becomes of the order of the
string scale, and the lowest order effective action~(\ref{Seff}) does
not give anymore a good description. We are in a full 'stringy'
regime. One expects that higher order corrections to the effective
action tame the growth of the curvature, and  both $(1/a)da/dt$ and
$d\phi/dt$ stay approximately constant. In terms of conformal time,
this means
\be\label{str}
a(\eta )=-\frac{1}{H_s\eta}\, ,\hspace{8mm}
\phi (\eta )=\phi_s-2\beta\log\frac{\eta}{\eta_s}
\ee
The stringy  phase lasts  for $\eta_s<\eta <\eta_1 <0$.
One then expects
that at this stage the dilaton potential becomes operative and, either
with a modification of the classical equations of motion
due to the dilaton potential~\cite{BV}, or
via quantum tunneling~\cite{GMV} the solution joins the standard
radiation dominated solution with constant dilaton, 
which is also a solution of the string
equations of motion derived from the action~(\ref{Seff}), with
external bulk stringy matter~\cite{GV}.
This gives, for $\eta_1 <\eta <\eta_r$, (with $\eta_r>0$)
\be\label{RD}
a(\eta )=\frac{1}{H_s\eta_1^2}\,(\eta -2\eta_1)\, ,\hspace{8mm}
\phi =\phi_0\, .
\ee
After that, the standard matter dominated era takes place.
We have chosen the additive and multiplicative
constants in $a(\eta )$ in such a way that 
$a(\eta)$ and $da/d\eta$ (and therefore also $da/dt$)
are continuous across the transitions.

The equation for the Fourier modes of
metric tensor perturbations for the two physical polarizations 
in the transverse traceless gauge is~\cite{GG}
\be\label{psi}
\frac{d^2\psi_k}{d\eta^2}+\left[ k^2-V(\eta )\right]\psi_k =0\, ,
\ee
\be\label{pot}
V(\eta )=\frac{1}{a}\,e^{\phi /2}\,\frac{d^2}{d\eta^2}(a\,
e^{-\phi /2})\, .
\ee
Inserting the expressions~(\ref{dil}-\ref{RD}) the potential is
\bees
&&V(\eta )= \frac{1}{4}\, (4\nu^2-1)\left(\eta-(1-\alpha)\eta_s
\right)^{-2},
\hspace*{10mm} -\infty <\eta <\eta_s \nonumber \\
&&V(\eta )= \frac{1}{4}\, (4\mu^2-1)\eta^{-2},\hspace*{40mm}
\eta_s <\eta<\eta_1\\\vr
&&V(\eta )= 0\,,\hspace*{64mm}\eta_1 <\eta<\eta_r \nonumber 
\ees
where $2\mu =|2\beta -3|,2\nu =|2\alpha-\gamma+1|$. The exact
solutions of eq.~(\ref{psi}) in the three regions are
\bees\label{sol}
&& \psi_k(\eta )=\sqrt{|\eta -(1-\alpha )\eta_s|}\, C
\,H_{\nu}^{(2)}(k|\eta -(1-\alpha )\eta_s|)\,,
\hspace*{2mm}-\infty <\eta<\eta_s\nonumber \\\vrbig
&& \psi_k(\eta )=\sqrt{|\eta|}\,\left[ A_+ \,H_{\mu}^{(2)}(k|\eta|)
+A_- \,H_{\mu}^{(1)}(k|\eta|)\right]\,,
\hspace*{13mm}\eta_s <\eta<\eta_1\\
&& \psi_k(\eta )=i\,\sqrt{\frac{2}{\pi k}}\,\left[ B_+\,e^{ik\eta}
-B_-\, e^{-ik\eta}\right]\,, \hspace*{30mm}\eta_1 <\eta<\eta_r 
\nonumber 
\ees
where $H_{\nu}^{(1,2)}$ are Hankel's functions.
The constants $A_{\pm},B_{\pm}$ can be obtained requiring the
continuity of the solution and of its derivative. We have chosen the
boundary conditions so that, at $\eta\ra -\infty$, $\psi_k\sim
\exp( ik\eta)$. In this case the number of particles created per
unit cell of the phase space is given by $|B_-|^2$. 

Before performing the matching, let us discuss the parameters of 
the model.
The two constants $\alpha,\gamma$ parametrize the 
solution in the dilaton dominated phase
and therefore they are fixed by the effective action~(\ref{Seff}):
$\alpha =1/(1+\sqrt{3}),\gamma =\sqrt{3}$ and then
 $\nu =0$ (anyway, we will write many of our results for generic $\nu$). 
Instead $\mu$ (or $\beta$) is a
free parameter which measures the growth of the dilaton during the
stringy phase; by definition $\mu \ge 0$.
The parameter $H_s$ is the Hubble constant during
the stringy phase. Since in this model the growth of the curvature 
can only be
stopped by the inclusion of higher order terms in the string
effective action, it is clear that the  natural value for $H_s$ is
of the order of the inverse of the string length $\lambda_s$.
If one uses  the value
$\lambda_s^2 \simeq (2/\alpha_{GUT})\,L_{\rm Pl}^2
\simeq 40\,L_{\rm Pl}^2$ then the typical value of $H_s$ is
$H_s\simeq 1/\lambda_s\simeq 0.15\,M_{\rm pl}$ where
$M_{\rm pl}$ is the Planck mass.
Finally, there are the two parameters $\eta_s,\eta_1$. In the
solution for $\psi_k$, and therefore in the spectrum, they appear in
the combinations $k|\eta_s|,k|\eta_1|$, where $k$ is the comoving wave
number. If we denote by $2\pi f$ the physical frequency observed at
a detector, we have $2\pi f=k/a(t_{\rm pres})$, where $t_{\rm pres}$
is the present value of cosmic time. Therefore, using eq.~(\ref{str}),
\be\label{eta1}
k|\eta_1|=2\pi fa(t_{\rm pres})|\eta_1|=
\frac{2\pi f}{H_s}\, \frac{a(t_{\rm pres})}{a(t_1)}=
\frac{2\pi f}{H_s}\left (\frac{t_{\rm pres}}{t_{\rm eq}}\right )^{2/3}
\left (\frac{t_{\rm eq}}{t_1}\right )^{1/2}\, ,
\ee
where $t_{\rm eq}\simeq 3.4\cdot 10^{10}\,h_0^{-4}s$ is the time of
matter-radiation equilibrium, and $t_{\rm pres}
=2/(3H_0)\simeq 2.1\cdot 10^{17}\,h_0^{-1}s$. 
The constant $h_0$ parametrizes the uncertainty in  the
present value of the Hubble constant 
$H_0= 3.2\cdot 10^{-18}\,h_0\,{\rm Hz}$,
and it cancels in
eq.~(\ref{eta1}); $t_1$ is the value of cosmic time  when the string
phase ends. In this context, the natural choice is
$t_1\simeq\lambda_s$.
Therefore the parameter $\eta_1$ can be traded for a
parameter $f_1$ defined by
\be
k|\eta_1|=\frac{f}{f_1}\, ,\hspace{10mm}f_1\simeq 
4.3\cdot 10^{10} {\rm Hz}\, \,
\left(\frac{H_s}{0.15\, M_{\rm pl}}\right)\,
\left(\frac{t_1}{\lambda_s}\right)^{1/2}
\, .
\ee
The order of magnitude of $f_1$
 is therefore  fixed. Note that at the frequencies of interest
for LIGO and VIRGO $f$ ranges  between 
$10\,{\rm Hz}$ and $1\,\rm{kHz}$, and $f/f_1$ is a very small quantity.
Similarly we can introduce a parameter $f_s$ instead of $\eta_s$,
from $k|\eta_s|={f}/{f_s}$. This parameter depends on the duration
of the string phase, and it is therefore totally unknown, even as an
order of magnitude. However, since $|\eta_1|<|\eta_s|$, we have
$f_s<f_1$.

To summarize, the model has a dimensionful parameter $f_s$, which
can have any value in the range
$0<f_s <f_1$
and a dimensionless parameter $\mu \ge 0$
(or equivalently $\beta$ with $2\mu=|2\beta -3|$). The
dimensionless constants $\alpha ,\nu$ are fixed,
$\alpha =1/(1+\sqrt{3}),\nu =0$ and the dimensionful
 constants $H_s,f_1$ are fixed within an uncertainty of about
one or two orders of magnitude. The constant $\phi_s$ appearing in
eq.~(\ref{dil}) drops out from the potential, eq.~(\ref{pot}),
and it is therefore irrelevant for our 
purposes.\footnote{For comparison,
in ref.~\cite{BGGV}  the two parameters which are not fixed 
are chosen as
$g_s/g_1$, which in our notations is $(f_s/f_1)^{\beta}$, and
$z_s =f_1/f_s$.}

\section{The spectrum}

Performing the matching at $\eta =\eta_s$ we get
\bees
A_{\pm}=&\pm & i\,\frac{\pi x_s\sqrt{\alpha}}{4}\,
\left[
{H_{\nu}^{(2)}}'(\alpha x_s)\,H_{\mu}^{(1,2)}(x_s)-
H_{\nu}^{(2)}(\alpha x_s)\,{H_{\mu}^{(1,2)}}'(x_s)\right.\nonumber\\
&+&\left.\frac{1}{2x_s}\,\frac{1-\alpha}{\alpha}\,H_{\nu}^{(2)}
(\alpha x_s)
\,H_{\mu}^{(1,2)}(x_s)\right]\, ,
\ees
where $x_s=f/f_s$; in $H_{\mu}^{(1,2)}(x_s)$, $H_{\mu}^{(1)}(x_s)$
refers to $A_+$ and $H_{\mu}^{(2)}(x_s)$ refers to
 $A_-$. In deriving these expressions we
have used the identity between Hankel functions
${H_{\nu}^{(2)}}'(x)\,H_{\nu}^{(1)}(x)-{H_{\nu}^{(1)}}'(x)\,
H_{\nu}^{(2)}(x)=
-4i/(\pi x)$. The constant $C$ appearing in eq.~(\ref{sol})
has been fixed requiring 
$|A_+|^2-|A_-|^2=1$, which gives $|C|=1$.
Next we perform the matching at $\eta =\eta_1$. However,  at the
frequencies of interest for LIGO and VIRGO,
$k|\eta_1|=f/f_1=O(10^{-8})$, and therefore in this second matching
we can use the
small argument limit of the Hankel functions, with a 
totally neglegible error. This 
gives a relatively simple analytical expression for the coefficient
$B_-$  which, apart from an irrelevant overall phase, is,
for $\mu\neq 0$,\footnote{For $\mu =0$ the small argument limit of the
Hankel function is different. The result for $\mu =0$ is the same
as  eq.~(\ref{b-}) if one writes
$2\log f/f_1$ instead of $\Gamma (\mu )$ and sets $\mu =0$ in
the remaining expression. In the following we write our formulae for
$\mu\neq 0$.}
\bees
& &B_-=\sqrt{\pi\alpha}\,\frac{2\mu -1}{8}\,\Gamma (\mu )\,
\left(\frac{f}{2f_s}\right)\,\left(\frac{f}{2f_1}\right)^{-\mu-1/2}
\left[
{H_{\nu}^{(2)}}'\left(\frac{\alpha f}{f_s}\right)\,
J_{\mu}\left(\frac{f}{f_s}\right)\right. \nonumber\\
\label{b-}
& &\left. -H_{\nu}^{(2)}\left(\frac{\alpha f}{f_s}\right)
\,J_{\mu}'\left(\frac{f}{f_s}\right) +
\frac{(1-\alpha)}{2\alpha}\,\frac{f_s}{f}\,
H_{\nu}^{(2)}\left(\frac{\alpha f}{f_s}\right)\,
J_{\mu}\left(\frac{f}{f_s}\right)\right ]\,,
\ees
where $J_{\mu}(z)$ is the Bessel function.
The spectrum of gravitational waves is expressed with the quantity
\be
\Omega_{\rm gw}(f)=\frac{1}{\rho_c}\,\frac{d\rho_{\rm gw}}{d\log f}=
\frac{1}{\rho_c}16\pi^2\,f^4\,|B_-|^2\,
\ee
where $\rho_c =3H_0^2M_{\rm pl}^2/(8\pi)$.
Then our result for the spectrum is
\bees
& & \Omega_{\rm gw}(f)=a(\mu )\,\frac{(2\pi f_s)^4}{H_0^2M_{\rm pl}^2}\,
\left(\frac{f_1}{f_s}\right)^{2\mu +1}\,
\left(\frac{f}{f_s}\right)^{5-2\mu}\,
 \left| 
{H_{\nu}^{(2)}}'\left(\frac{\alpha f}{f_s}\right)
\,J_{\mu}\left(\frac{f}{f_s}\right) \right.\nonumber\\
& & \left. -H_{\nu}^{(2)}\left(\frac{\alpha f}{f_s}\right)
\,J_{\mu}'\left(\frac{f}{f_s}\right) 
+\frac{(1-\alpha)}{2\alpha}\,\frac{f_s}{f}\,
H_{\nu}^{(2)}\left(\frac{\alpha f}{f_s}\right)\,
J_{\mu}\left(\frac{f}{f_s}\right)
\right|^2
\ees
where
\be
a(\mu )= \frac{\alpha}{48}\, 2^{2\mu}\,(2\mu -1)^2\,\Gamma^2(\mu )\, .
\nonumber
\ee
In the most interesting case $\nu =0$, using the identity
${H_0^{(2)}}'(z)=-H_1^{(2)}(z)$, we can rewrite the spectrum as
\bees\label{res}
& & \Omega_{\rm gw}(f)=a(\mu )\,\frac{(2\pi f_s)^4}{H_0^2M_{\rm pl}^2}\,
\left(\frac{f_1}{f_s}\right)^{2\mu +1}\,
\left(\frac{f}{f_s}\right)^{5-2\mu}\,
 \left| H_0^{(2)}\left(\frac{\alpha f}{f_s}\right)\,
J_{\mu}'\left(\frac{f}{f_s}\right)\right. +\nonumber\\
& &\left. +H_1^{(2)}\left(\frac{\alpha f}{f_s}\right)
\,J_{\mu}\left(\frac{f}{f_s}\right)
 -\frac{(1-\alpha)}{2\alpha}\,\frac{f_s}{f}\,
H_{0}^{(2)}\left(\frac{\alpha f}{f_s}\right)\,
J_{\mu}\left(\frac{f}{f_s}\right)
\right|^2\, .
\ees
Expanding our exact expression for small values of $f/f_s$ we get,
for $\nu =0$,
\bees\label{low}
& &\Omega_{\rm gw}(f)\simeq \frac{(2\mu -1)^2}{192\mu^2\alpha}\,
\frac{(2\pi f_s)^4}{H_0^2M_{\rm pl}^2}\left(\frac{f_1}{f_s}
\right)^{2\mu +1}
\left(\frac{f}{f_s}\right)^3\,\times\nonumber \\
& &
\left\{  (2\mu\alpha -1+\alpha)^2 
+\frac{4}{\pi^2}\left[
(2\mu\alpha -1+\alpha)\,\left (\log\frac{\alpha f}{2f_s}+\gamma_E
\right )-2
\right]^2\right\}\, ,
\ees
where $\gamma_E\simeq 0.5772\ldots$ is Euler constant.
This expression agrees  with the result obtained in the literature,
see eq.~(5.7) of ref.~\cite{review2}, apart for the numerical
constants which cannot be computed using only the approximate solution
discussed in refs.~\cite{review2,BGGV}.

Let us observe that the low frequency limit in which eq.~(\ref{low})
holds is actually $f\ll f_s\ll f_1$. If we are interested in the limit 
$f\ll f_s\sim f_1$ we should not take the small argument limit 
of the Hankel function when performing  the matching at $\eta_1$. 
Rather, we must keep the exact expression and perform the expansion in
the final result. If we do not make any assumption on the value of
$f_1/f_s$  a straigthforward computation shows that 
in the  limit $f/f_s\ll 1,f/f_1\ll 1$
\bees
B_-\simeq \frac{1}{4\mu\sqrt{\pi\alpha}}\times\hspace*{80mm}& &
\nonumber\\
\left\{ \left(\mu-\frac{1}{2}\right )
\left[ 1+\frac{i\pi}{4}(1-\alpha-2\mu\alpha )\left(1-\frac{2i}{\pi}
\left(\log \frac{\alpha f}{2f_s}+\gamma_E\right)\right)\right]
\left(\frac{f}{2f_s}\right)^{\mu}\left(\frac{f}{2f_1}
\right)^{-\mu -1/2}\right.& &\nonumber\\
\left. +\left(\mu+\frac{1}{2}\right)
\left[ 1+\frac{i\pi}{4}(1-\alpha+2\mu\alpha )\left(1-\frac{2i}{\pi}
\left(\log \frac{\alpha f}{2f_s}+\gamma_E\right)\right)\right]
\left(\frac{f}{2f_s}\right)^{-\mu}\left(\frac{f}{2f_1}
\right)^{\mu -1/2}\right\}& & \nonumber
\ees
from which we derive again eq.~(\ref{low}) if we now take $f_s\ll
f_1$. As we will see below, the graviton spectrum is
neglegibly small unless $f_s\ll f_1$, so the physically relevant limit
is the one which leads to
 eq.~(\ref{low}). If we instead consider the spectrum
with $\nu >0$ a simple calculation gives a 
low frequency behavior $\sim f^{3-2\nu}$, without logarithmic
corrections (the absence of the $\log f$ term  is due to the different
small argument limit of $H_{\nu}^{(1,2)}$ for $\nu =0$ and for $\nu >0$.)

Expanding eq.~(\ref{res})
in the  limit $f\gg f_s$ (but still $f \ll f_1$ since  eq.~(\ref{res})
holds only in this limit) we find instead
\bees
\Omega_{\rm gw}(f)&\simeq &\frac{4a(\mu )}{\pi^2\alpha}\,
\frac{(2\pi f_s)^4}{H_0^2M_{\rm pl}^2}
\,\left(\frac{f_1}{f_s}\right)^{2\mu +1}
\,\left(\frac{f}{f_s}\right)^{3-2\mu}\nonumber\\
\label{large}
 &=&\frac{4a(\mu )}{\pi^2\alpha}\,
\frac{(2\pi f_1)^4}{H_0^2M_{\rm pl}^2}\,
\left(\frac{f}{f_1}\right)^{3-2\mu}
\ees
which  agrees, in the frequency dependence, with the result of
ref.~\cite{BGGV}.\footnote{For the comparison with ref.~\cite{BGGV},
note that, since $2\mu =|2\beta -3|$, if $2\beta >3$ the dependence on
$f$ is $\sim f^{3-2\mu}=f^{6 -2\beta}$ while, if $2\beta <3$,
$f^{3-2\mu}=f^{2\beta}$, which therefore reproduces eq.~(3.5) of 
ref.~\cite{BGGV}.} It is important to stress that in the high
frequency limit the unknown parameter $f_s$  cancels.

Finally, at sufficiently large  $f$, there is a rather sharp cutoff
and the spectrum goes to zero exponentially. The cutoff can be
obtained computing the spectrum without performing the limit
$f\ll f_1$ in the second matching. More simply, the cutoff
frequency $f_{\rm max}$ can be estimated
from $k_{\rm max}^2\simeq |V(\eta_1)|$, which gives 
\be
f_{\rm max}\simeq\frac{1}{2}\,\sqrt{|4\mu^2-1|}\, f_1\, .
\ee

\section{Discussion}

{}From eq.~(\ref{large}) we see that the form of the spectrum depends
crucially on whether $\mu <3/2$, $\mu =3/2$ or $\mu >3/2$. Let us
consider first the case $\mu >3/2$. In this case the spectrum is
a decreasing function of $f$ if $f\gg f_s$. Numerically,
\be\label{scala}
\frac{(2\pi f_1)^4}{H_0^2M_{\rm pl}^2}\simeq 1.4\cdot 10^{-6}\,
\frac{1}{h_0^2}\, \left(\frac{H_s}{0.15\,M_{\rm pl}}\right)^4\,
\left(\frac{t_1}{\lambda_s}\right)^2 \, ,
\ee
and in eq.~(\ref{large}) this number is multiplied by 
$(f_1/f)^{2\mu -3}$; $f_1/f$
is $O(10^8)$ at $f=100\,{\rm Hz}$ 
and even larger for smaller frequencies, while
$2\mu -3$ is positive in this case. Therefore, for
$\mu >3/2$, $\Omega_{\rm gw}$ would violate any experimental bound. More
precisely,  the computation becomes invalid because we should include
the back-reaction of the produced gravitons on the
metric~\cite{BGGV}. We will therefore consider only $0< \mu\leq 3/2$.
In this case, the spectrum  at low frequencies  increases as $\sim
f^3\log^2 f$, and at high frequencies is increasing as
$f^{3-2\mu}$ (or going to a constant if $\mu =3/2$.)

Fig.~1  shows the form of the spectrum for $\mu =1.4$ and for $\mu
=3/2$. In this figure we plot $\Omega_{\rm gw}(f)$,
measured in units of 
$a(\mu ) \left[ (2\pi f_s)^4/(H_0^2M_{\rm pl}^2)\right]\,
(f_1/f_s)^{2\mu +1}$, which is
the overall constant appearing in
eq.~(\ref{res}), versus $f/f_s$. We see that, compared to the low- and
high-frequency expansions discussed in~\cite{BGGV}, the spectrum also
displays a series of oscillations. Depending on the value of $f_s$,
the window available to the LIGO and VIRGO interferometers, 
$f$ between $10\,{\rm Hz}$ and $1\,{\rm kHz}$, may contain 
many oscillations, and this
would provide a rather characteristic signature.

To give an idea of the magnitude of the spectrum, in
figs.~2 and~3 we plot
$h_0^2\Omega (f)$ for $\mu =1.4$ and for $\mu =3/2$,  
for a specific value
of $f_s$,  $f_s=100\,{\rm Hz}$ 
(this choice of parameters is motivated below), 
and for $H_s=0.15M_{\rm pl}$ and $t_1=\lambda_s$, in the
frequency range relevant for LIGO and VIRGO.
(Note that the quantity of interest for the experimentalist is not 
$\Omega_{\rm gw}$ but $h_0^2\,\Omega_{\rm gw}$, since $h_0$ 
only reflects
our uncertainty in the quantity which we use
to normalize the result.)

It is also useful to give the result in terms of the quantity
$h_c(f)$, which is the dimensionless strain $\Delta L/L$ produced in
the arms of the detector, and
is related to $\Omega_{\rm gw}(f)$ by~\cite{Th}
\be
h_c(f)\simeq 1.3\cdot 10^{-20}\,\sqrt{h_0^2\,\Omega_{\rm
gw}(f)}\,\left (\frac{100\,{\rm Hz}}{f}\right) \, .
\ee
Fig.~4 shows a plot of $h_c(f)$ versus $f$ for 
$f_s=10$ Hz and for $\mu =3/2$ and $\mu =1.4$.

Let us then discuss what is the best possible result that we 
can obtain
from this model, varying the two parameters $f_s$ and $\mu$, with 
$0<f_s<f_1$ and $0<\mu\leq 3/2$. Suppose that we want to
detect a signal at a given frequency, say $f= 100\,{\rm Hz}$. 
In fig.~5 we
plot $\Omega_{\rm gw}$, from eq.~(\ref{res}),
 as a function of $f_s$ at fixed $f$. We see that,
independently of $\mu$, it increases for decreasing $f_s$, and
when $f_s\ll f$ it
reaches asymptotically the constant value given by eq.~(\ref{large}).
So, if we want to detect a signal at a given frequency
$f$, the optimal situation is obtained 
if the value of $f_s$ is smaller than
$f$. How much smaller is not 
very important, since as a function of $f_s$,
$\Omega_{\rm gw}$ saturates and practically
reaches its maximum value as soon as, say,
$f_s<0.5 f$. The maximum $\Omega_{\rm gw}$ is therefore given by
eq.~(\ref{large}), which still depends on the other parameter
$\mu$. Since $f/f_1$ is a very small parameter, we see immediately
that the best possible situation is realized when $\mu =3/2$. Note
that this means $\beta =0$ or $\beta =3$, 
and   in the first case
not only the  derivative of the dilaton with respect to cosmic time,
but even the dilaton itself stays constant during the stringy phase.

In this case, $\Omega_{\rm gw}$ reaches a maximum value
\be\label{max}
h_0^2\,\Omega_{\rm gw}^{\rm max}=
\frac{2h_0^2}{3\pi}\,\frac{(2\pi f_1)^4}{H_0^2M_{\rm pl}^2}
\simeq 3.0\cdot 10^{-7}\, \left(\frac{H_s}{0.15\,M_{\rm pl}}\right)^4\,
\left(\frac{t_1}{\lambda_s}\right)^2\, .
\ee

If $\mu =3/2$, this maximum
value is reached as long as $f>f_s$ and therefore, if $f_s$ is smaller
than, say, $10\,{\rm  Hz}$, it is already
reached in the VIRGO/LIGO frequency range; after that, the signal
oscillates around a constant value (fig.~3). If
instead $\mu <3/2$, eq.~(\ref{large}) shows that there is a further
suppression factor $(f/f_1)^{3-2\mu}$, and therefore the maximum value
is reached only at the cutoff frequency
$f_{\rm max}\sim f_1$, 
that is for frequencies  around $10$ or $100\,\rm{GHz}$. 
However,  if $\mu$ is not close enough to 3/2, the suppression factor 
$(f/f_1)^{3-2\mu}$ makes the signal very small at LIGO/VIRGO
frequencies, 
unless one uses unnaturally large values of $H_s,t_1$ (fig.~2).

Let us then discuss  whether
this maximum value is compatible with the
experimental constraints mentioned in the Introduction.

We consider first the nucleosynthesis bound~\cite{Allenrev,peak}
\be
\int \Omega_{\rm gw}(f)\,d(\log f)\leq 
\frac{7}{43}(N_{\nu}-3) \left( 1+z_{\rm eq}\right)^{-1}\, ,
\ee
where $1+z_{\rm eq}\simeq 2.32\times 10^{4}h_0^2$ is the redshift at
the time of radiation-matter equilibrium and $N_{\nu}$ is the
equivalent number of neutrino species~\cite{KT}. Using the recent
analysis  of ref.~\cite{Copi}, $N_{\nu}<3.9$, we get
\be
\int h_0^2\,\Omega_{\rm gw}(f)\,d(\log f) < 6.3 \cdot 10^{-6}\, .
\ee
For $\mu =3/2$, $\Omega_{\rm gw}$ reaches its maximum value at $f\sim
f_s$ and stays approximately constant untill the cutoff at $f\sim f_1$
is reached, and therefore the bound gives
\be\label{nsbound}
h_0^2\,\Omega_{\rm gw}^{\rm max}\,\log\frac{f_1}{f_s}\stackrel{<}{\sim}
6.3 \cdot 10^{-6}\, ,
\ee
If we take  $f_s\sim 100\,\rm{Hz}$ (which still satisfies 
$f_s <f$ at the frequency $f=1\,{\rm kHz}$ accessible to LIGO 
and VIRGO, so
that eq.~(\ref{max}) applies), eq.~(\ref{nsbound}) gives an upper
bound on the spectrum predicted by string cosmology
\be\label{max2}
h_0^2\,\Omega_{\rm gw}^{\rm max}<
3.2\cdot 10^{-7}\, ,
\ee
which can be obtained from eq.~(\ref{max}) with very natural values of
$H_s,t_1$. This number
is smaller than the planned  sensitivity of 
the LIGO and VIRGO detectors in a first phase but well within the
planned sensitivity of the advanced project.
Smaller values of $f_s$ give a more stringent bound, but the
dependence is only logarithmic: if $f_s\sim 10^{-7}\,{\rm Hz}$, 
the maximum
value of $h_0^2\Omega_{\rm gw}$ is $1.5\cdot 10^{-7}$.

The  bound from $\mu$sec pulsars is~\cite{Allenrev} 
\be
h_0^2\,\Omega_{\rm gw}(f=10^{-8}{\rm Hz})<10^{-8}\, .
\ee
{}From fig.~2 we see that, in order to suppress the result at
$f=10^{-8}{\rm Hz}$, we must have $f_s\gg 10^{-8}{\rm Hz}$, which
is well compatible with the condition
$f_s\lsim 10^2\,{\rm Hz}$ required before.
Even if at $f\sim1\,{\rm kHz}$ we have
$h_0^2\,\Omega_{\rm gw}^{\rm max} = 3.2\cdot 10^{-7}$, 
at $f\ll f_s$ we get
\be
h_0^2\,\Omega_{\rm gw}\simeq 3.3\cdot 10^{-9}\,
\left(\frac{f}{f_s}\right)^3\,\log^2\frac{f}{f_s}\, .
\ee
We see that
a value of, say, $f_s> 10^{-7}\,{\rm Hz}$ is sufficient to bring
$\Omega_{\rm gw}(f=10^{-8}{\rm Hz})$ well  below the
experimental bound. The COBE bound is even more easily satisfied since
if $f_s> 10^{-7}\,{\rm Hz}$ the 
value of $\Omega_{\rm gw}$ at $f =10^{-16}\,{\rm Hz}$ is totally
neglegible. Fig.~6 shows the spectrum for
$f_s =100\,{\rm Hz}$ and $\mu =3/2$ in a large range of frequencies, and
compares it to the experimental bounds.

In conclusion, in the most favourable case
$\mu =3/2$ and $10^{-7} {\rm Hz}< f_s \lsim 100\,{\rm Hz}$, 
the relic gravitational waves background predicted by string
cosmology  at the frequencies of LIGO and VIRGO is about
$h_0^2\,\Omega_{\rm gw}^{\rm max}=3.2\cdot 10^{-7}$,
which is smaller 
than the planned sensitivity for coincidence experiments 
with interferometers in the
first phase, but well within the sensitivity at which the advanced
LIGO project aims. This maximum value of $\Omega_{\rm gw}$ might also
be comparable to the sensitivities which could be reached
correlating resonant bar detectors such as EXPLORER, NAUTILUS and
AURIGA~\cite{Cerd}. 
At the same time, the spectrum satisfies the existing
experimental bounds.

\vspace{2mm}
{\bf Acknowledgements}  \newline
We are grateful to Maurizio Gasperini
and Gabriele Veneziano for very useful comments.

\newpage
\centerline{\bf Figure captions}
\begin{enumerate}
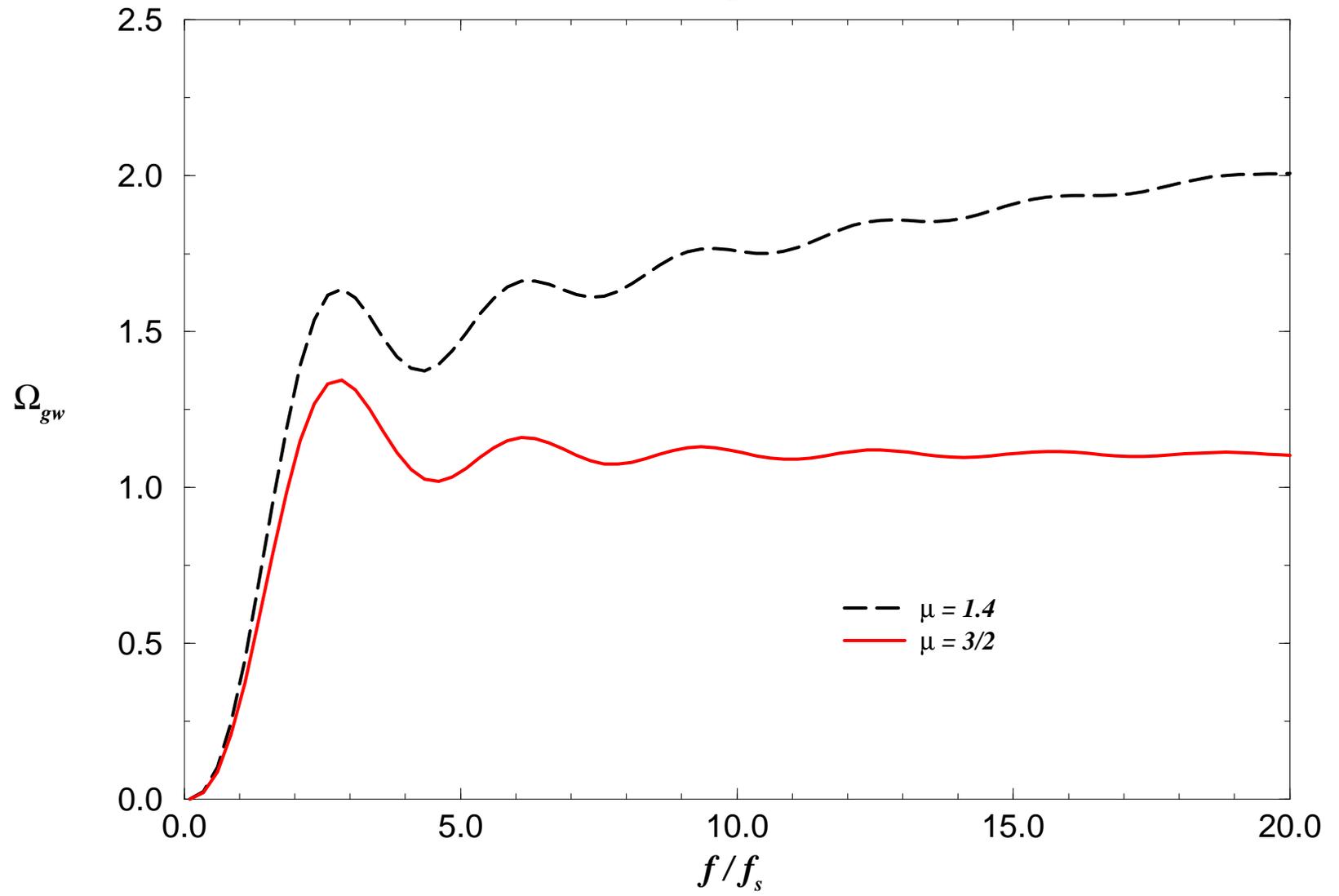
\item[Fig. 1] $\Omega_{\rm gw}$,
measured in units of $a(\mu ) ((2 \pi f_s)^4/(H_0^2\,M_{\rm pl}^2))
(f_1/f_s)^{2 \mu +1}$ vs. $f/f_s$ for $\mu=1.4$ and $\mu=1.5$.
\vskip 5mm
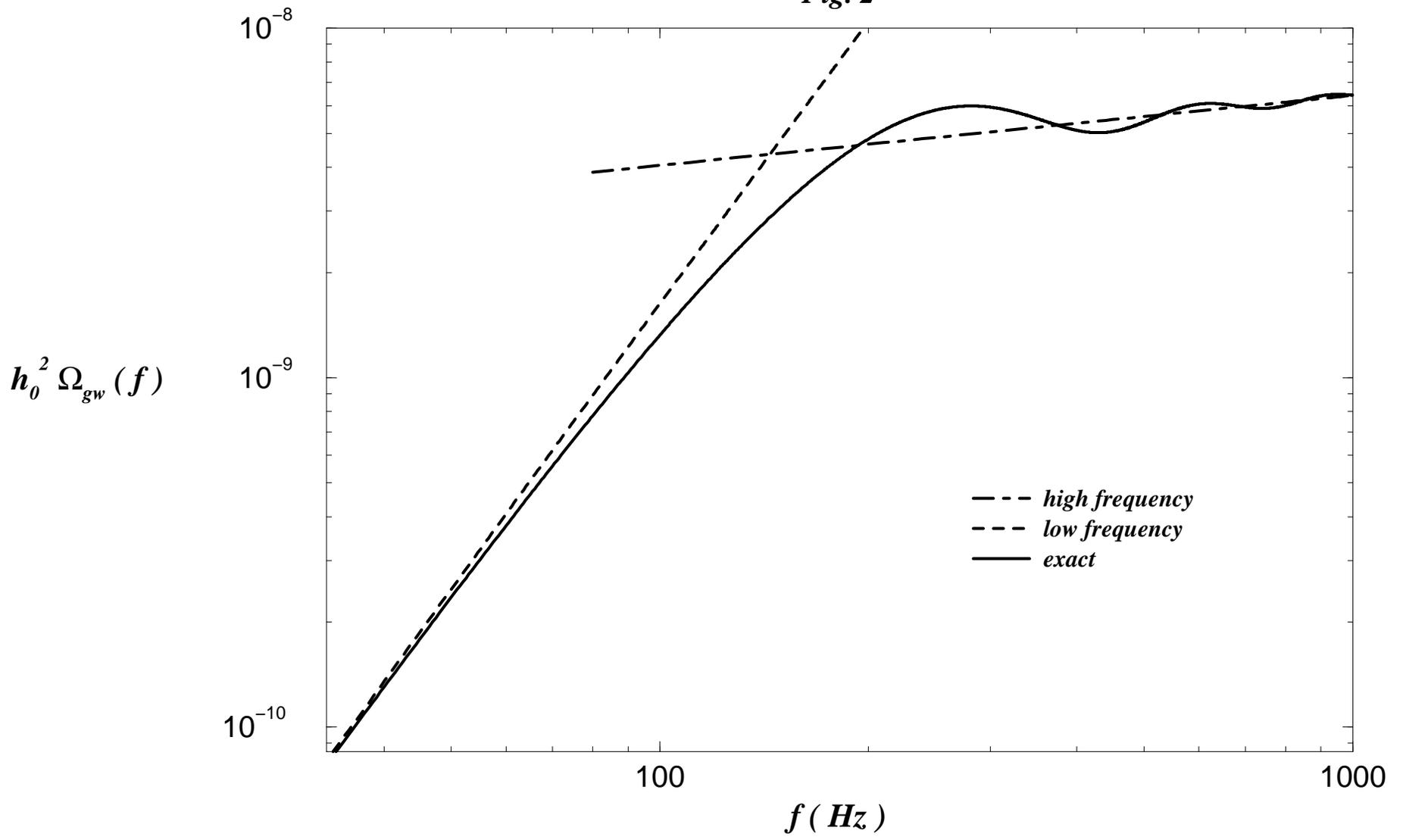
\item[Fig. 2]  $\Omega_{\rm gw}(f)$ vs. $f$ for $ \mu=1.4$,
$f_s=100\,{\rm Hz}$ and $f_1=4.3\cdot10^{10}\,{\rm Hz}$;
for comparison we also show the low- and high-frequencies limits.
\vskip 5mm
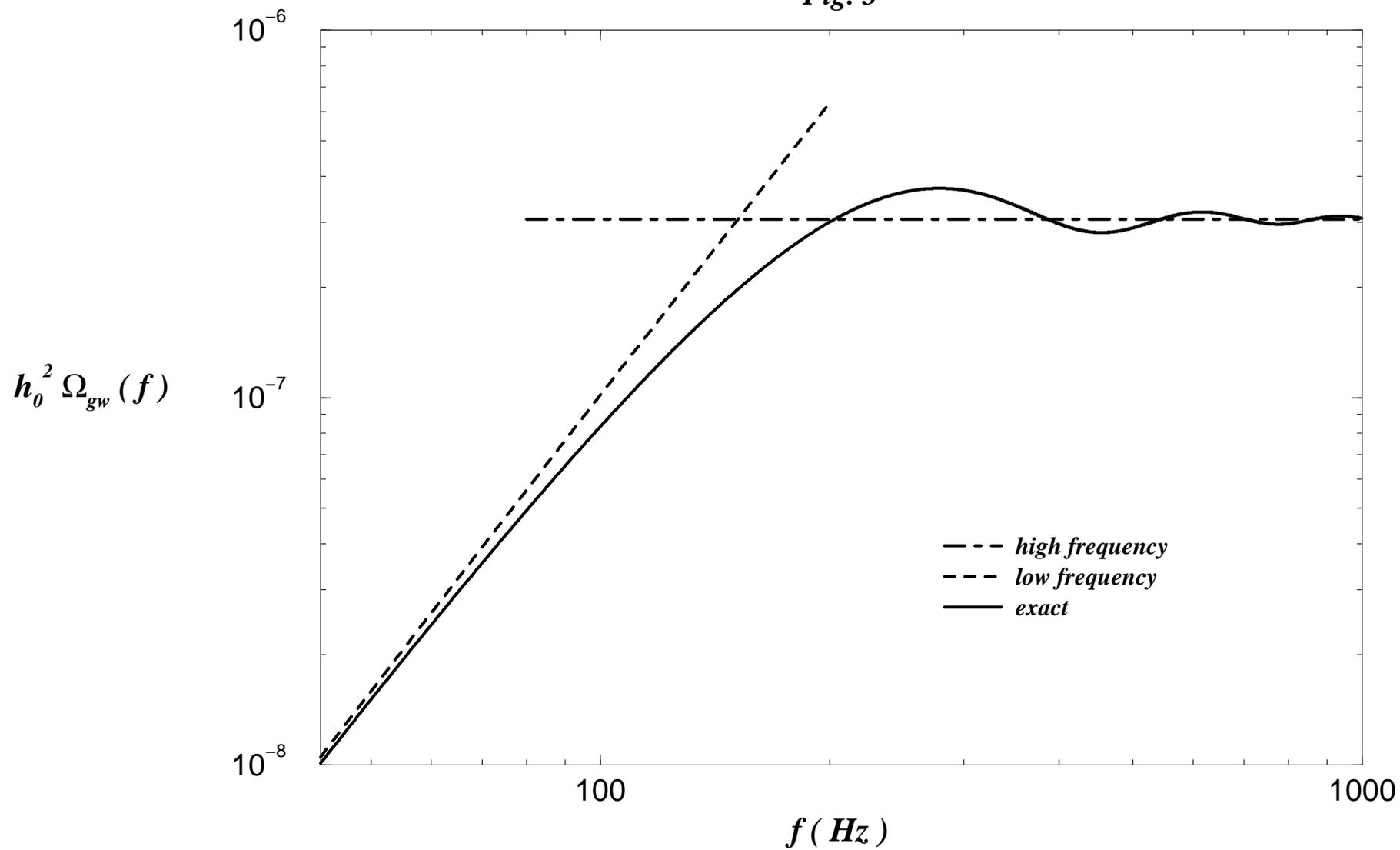
\item[Fig. 3]  $\Omega_{\rm gw}(f)$ vs. $f$ for $ \mu=1.5$,
$f_s=100\,{\rm Hz}$ and $f_1=4.3\cdot10^{10}\,{\rm Hz}$;
for comparison we also show the low- and high-frequencies limits.
\vskip 5mm
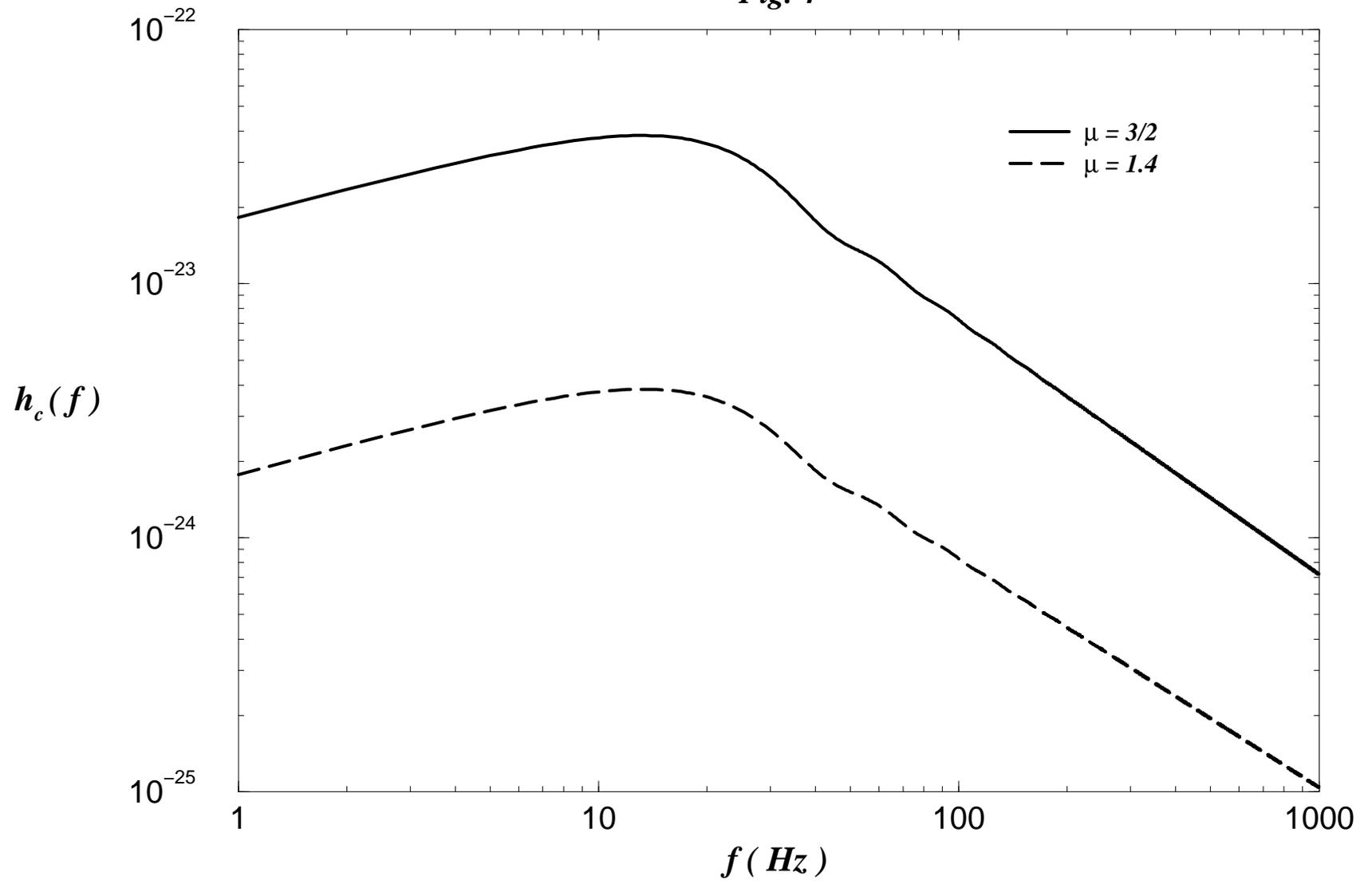
\item[Fig. 4]   $h_c(f)$ vs. $f$ for $ \mu=1.5$ and $\mu=1.4$,
$f_s=10\,{\rm Hz}$ and $f_1=4.3\cdot10^{10}\,{\rm Hz}$.
\vskip 5mm
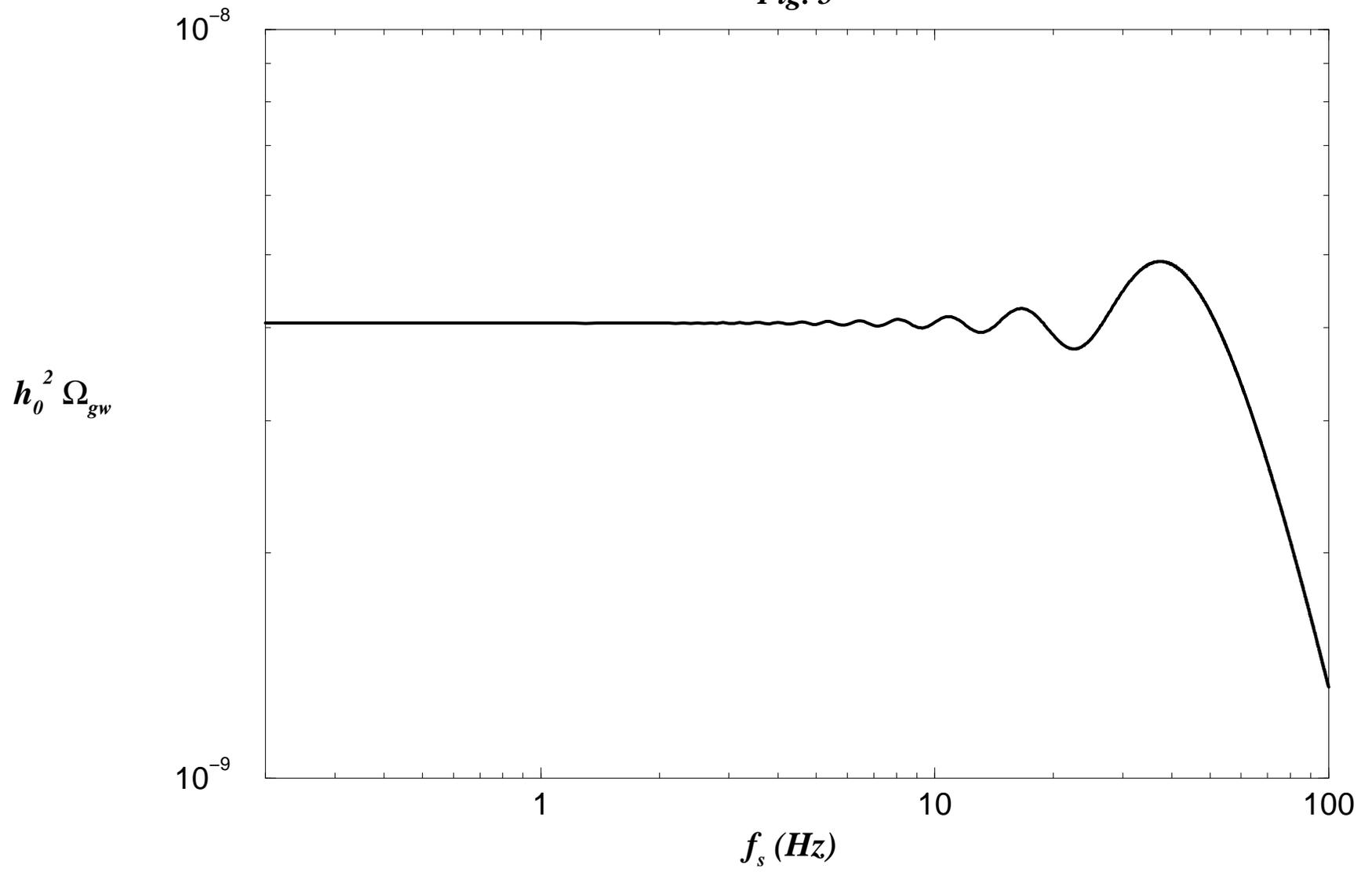
\item[Fig. 5]  $\Omega_{\rm gw}$ vs. $f_s$ for $ \mu=1.4$,
at fixed $f=100\,{\rm Hz}$ and $f_1=4.3\cdot10^{10}\,{\rm Hz}$.
\vskip 5mm
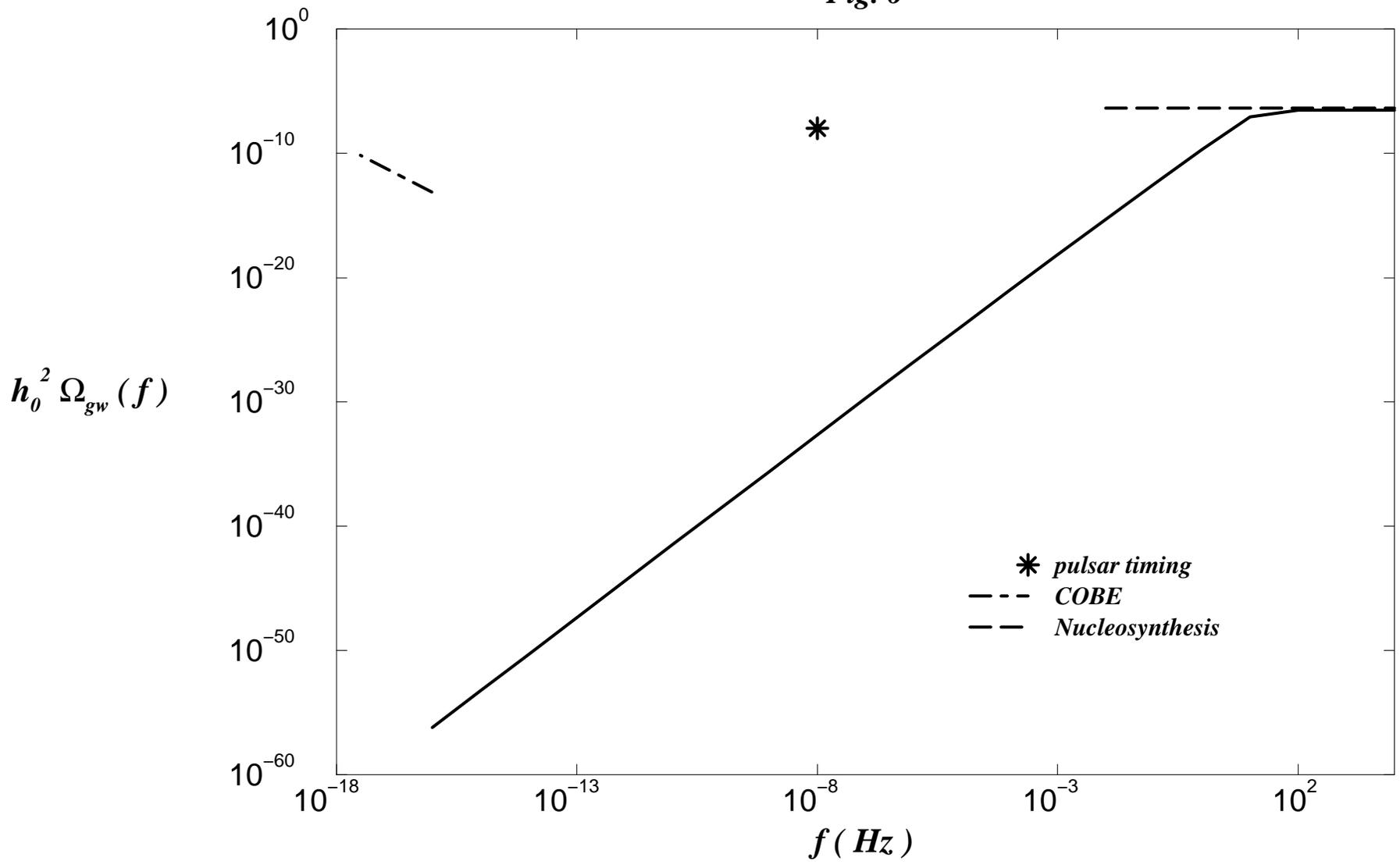
\item[Fig. 6]   $\Omega_{\rm gw}(f)$ vs. $f$ for $ \mu=1.5$,
$f_s=10\,{\rm Hz}$ and $f_1=4.3\cdot10^{10}\,{\rm Hz}$
compared to the experimental bounds.
\end{enumerate}

\newpage
\begin{thebibliography}{999}
\bibitem{Gri} L.P. Grishchuk, Sov. Phys. JETP, 40 (1975) 409.
\bibitem{vari} A. Starobinski, JETP Lett. 30 (1979) 682;

V. Rubakov, M. Sazhin and A. Veryaskin, Phys. Lett. 115B (1982) 189;

R. Fabbri and M.D. Pollock, Phys. Lett. 125B (1983) 445;

L. Abbott and M. Wise, Nucl. Phys. B244 (1984) 541.

L. Abbott and D. Harari, Nucl. Phys. B264 (1986) 487.

B. Allen, Phys. Rev. D37 (1988) 2078.
\bibitem{Gri2} L.P. Grishchuk, Class. Quantum Grav. 10 (1993) 2449.
\bibitem{Muk} V.F. Mukhanov, H.A. Feldman and R.H. Brandenberger,
Phys. Rep. 215 (1992) 203.
\bibitem{Allenrev} B. Allen, {\em ``The Stochastic Gravity-Wave
Background: Sources and Detection''}, gr-qc 9604033.
\bibitem{BGGV} R. Brustein, M. Gasperini, M.~Giovannini and
G.~Veneziano, Phys. Lett. B361 (1995) 45.
\bibitem{GV} M. Gasperini and G. Veneziano, Astropart. Phys. 1 (1993)
317; Mod. Phys. Lett. A8 (1993) 3701; Phys. Rev. D50 (1994) 2519.
\bibitem{review} G. Veneziano, {\em ``String Cosmology: Basic Ideas and
General Results''}, in {\em ``String Gravity and Physics at the Planck
Energy Scale''}, N.~Sanchez and A.~Zichichi eds.,
Kluwer Publ., pag 285.

\bibitem{review2}
M. Gasperini, {\em ``Status of String Cosmology: Phenomenological
Aspects''},  in {\em ``String Gravity and Physics at the Planck
Energy Scale''}, N.~Sanchez and A.~Zichichi eds.,
Kluwer Publ., pag 305.

\bibitem{BV} R. Brustein and G. Veneziano, Phys. Lett. B329 (1994) 429.
\bibitem{GMV} M. Gasperini, J. Maharana and G. Veneziano,
{\em ``Graceful exit in quantum string cosmology''}, Cern-Th/96-32;
hep-th/9602087, Nucl. Phys. B, in press.

M. Gasperini and G. Veneziano, {\em ``Birth of the Universe as quantum
scattering in string cosmology''}, Cern-Th/96-49,
hep-th/9602096, Gen. Rel. Grav., in press.
\bibitem{GG} M. Gasperini and M. Giovannini, Phys. Rev. D47 (1993) 1519.
\bibitem{Th} K. S. Thorne, in ``300 Years  of Gravitation'',
S. Hawking and W. Israel eds., Cambridge Univ. Press, 1987.
\bibitem{peak} R. Brustein, M. Gasperini and G. Veneziano, {\em ``Peak
and End Point of the Relic Graviton Background in String Cosmology''},
preprint CERN-TH/96-37, hep-th/9604084.
\bibitem{KT} E. Kolb and R. Turner, {\em ``The Early Universe''},
Addison-Wesley Publ., 1990
\bibitem{Copi} C. Copi et al., Phys. Rev. Lett. 75 (1995) 3981.
\bibitem{Cerd} M. Cerdonio et al., {\em ``Status of the Auriga
Gravitational Wave Antenna and Perspectives for the Gravitational
Waves Search with Ultracryogenic Resonant Mass Detectors''}, in
``Proc. of the First Edoardo Amaldi Conference'', ed. by E.~Coccia,
G.~Pizzella and F.~Ronga (World Scientific, Singapore, 1995).

P. Astone et al., {\em ``Upper limit for a gravitational wave
stochastic background measured with the EXPLORER and NAUTILUS
gravitational wave resonant detectors''}, (Rome, February 1996), to
appear. 
\end{thebibliography}
\end{document}